\documentclass[a4paper,11pt]{article}
\usepackage{pos}

\usepackage{subcaption}

\title{Investigating the capability of the Cherenkov Telescope Array Observatory to detect gamma-ray emission from simulated stationary neutrino sources identified by KM3NeT}
\ShortTitle{CTAO performance - KM3NeT}

\author*[a,b]{G.~M.~Cicciari}
\author[a,b]{M.~Mallamaci}
\author[a,b]{G.~Marsella}
\author[c]{A.~Rosales de Le\'{o}n}
\author[d,e,f]{O.~Sergijenko} 
\author[ ]{for the CTA Consortium}
\author[g]{G.~Ferrara}

\affiliation[a]{Università di Palermo, Dipartimento di Fisica e Chimica “E. Segrè”, Palermo, Italy}
\affiliation[b]{INFN Sezione di Catania, Catania, Italy}
\affiliation[g]{Università di Catania \& INFN Laboratori Nazionali del Sud, Catania, Italy}
\affiliation[c]{University of Lodz, Faculty of Physics and Applied Informatics, Department of Astrophysics, Lodz, Poland}
\affiliation[d]{Astronomical Observatory, Taras Shevchenko National University of Kyiv, Kyiv, Ukraine}
\affiliation[e]{Main Astronomical Observatory, National Academy of Sciences of Ukraine, Kyiv, Ukraine}
\affiliation[f]{AGH University of Science and Technology, Krakow, Poland}

\emailAdd{gloriamaria.cicciari@unipa.it}
\emailAdd{giovanna.ferrara@dfa.unict.it}
\emailAdd{manuela.mallamaci@unipa.it}
\emailAdd{giovanni.marsella@unipa.it}
\emailAdd{alberto.rosales@uni.lodz.pl}
\emailAdd{olga.sergijenko.astro@gmail.com}

\abstract{

The simultaneous observation of gamma rays and neutrinos from the same astrophysical source offers a unique opportunity to probe particle acceleration and interaction mechanisms in ultra-high-energy environments. The Cherenkov Telescope Array Observatory (CTAO) is a next-generation ground-based gamma-ray facility, sensitive to energies from 20 GeV to 300 TeV. In this work, we present for the first time a performance study of CTAO based on joint simulations of steady-state sources emitting both neutrinos and gamma rays, under the assumption that neutrino events are detected by the KM3NeT telescope in the Northern Hemisphere. To identify potentially observable sources, we apply a neutrino-based selection filter according to KM3NeT’s discovery potential. We then simulate gamma-ray detectability with CTAO, taking into account visibility, sensitivity, and extragalactic background light absorption.
The analysis is specifically focused on exploring the detectability of sources at low neutrino luminosities, limited to values below $10^{52} erg yr^{-1}$, in order to assess the performance of CTAO and KM3NeT in identifying faint extragalactic emitters. Particular attention is given to the strategic role of KM3NeT’s geographic location, which provides access to Southern-sky sources, and to the impact of the planned CTA+ upgrade, which will enhance CTAO-South with Large-Sized Telescopes (LSTs). Our results highlight the importance of coordinated multi-messenger strategies between KM3NeT and CTAO to maximize the discovery potential of astrophysical neutrino sources.

}

\ConferenceLogo{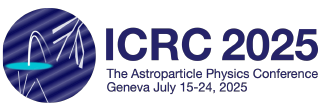}

\FullConference{39th International Cosmic Ray Conference (ICRC2025)\\
 15–24 July 2025\\
Geneva, Switzerland\\}

\begin{document}
\maketitle


\section{Introduction}
\label{Sec:Introduction}
\noindent
The identification of astrophysical neutrino sources remains a major goal in multi-messenger astronomy. High-energy neutrinos are unambiguous tracers of hadronic processes and provide a unique probe of extreme environments such as active galactic nuclei (AGNs), gamma-ray bursts, and starburst galaxies. However, their weak interactions and limited angular resolution make source association challenging. Gamma rays produced in hadronic interactions, typically via $\pi^0$ decay, can serve as electromagnetic counterparts to neutrino events. Joint gamma-ray and neutrino observations offer a powerful strategy for source identification and for testing models of cosmic ray acceleration. The Cherenkov Telescope Array Observatory (CTAO) and KM3NeT represent the next-generation facilities in gamma ray and neutrino astronomy, respectively. This study investigates their combined detection capability for steady extragalactic sources by simulating realistic populations and evaluating the gamma-ray detectability with CTAO under KM3NeT trigger assumptions. Particular emphasis is placed on the complementary sky coverage offered by KM3NeT's Northern Hemisphere location and the dual-site configuration of CTAO.

\section{Instruments and Framework}

\subsection{The Cherenkov Telescope Array Observatory (CTAO)}
\label{Sec:CTAO}
\noindent
CTAO is a ground-based gamma ray observatory under construction with two sites: CTAO-North in La Palma (Spain) and CTAO-South in Paranal (Chile). It will cover an energy range from 20~GeV to 300~TeV, combining Large-Sized Telescopes (LSTs), Medium-Sized Telescopes (MSTs), and Small-Sized Telescopes (SSTs) to optimize angular resolution, sensitivity, and sky coverage in its full energy domain~\cite{hofmann2024cherenkov}. The “Alpha” configuration considered in this work represents the initial deployment of telescopes and includes full MST coverage, but no LSTs at the southern site. This asymmetry has important implications for low-energy gamma-ray sensitivity in the Southern Hemisphere.

\subsection{The KM3NeT Neutrino Telescope}
\label{Sec:KM3NeT}
\noindent
KM3NeT ~\cite{km3netoverview} is a deep-sea neutrino telescope currently in deployment in the Mediterranean Sea. It consists of two detectors: ORCA for low-energy neutrinos and ARCA for high-energy astrophysical neutrino detection. The ARCA detector, located offshore from Sicily, is optimized for neutrinos above 10~TeV and has a subdegree angular resolution for track-like events~\cite{adrian2016letter}. Unlike IceCube ~\cite{icecube}, which is located at the southern pole, the position of KM3NeT in the northern hemisphere allows it to monitor a significant portion of the southern sky, particularly relevant for extragalactic source populations concentrated in this region.

\subsection{Multi-messenger Strategy}
\noindent 
The coordinated use of KM3NeT and CTAO opens new avenues for time-dependent and population studies. Neutrino events from KM3NeT, particularly those exceeding a 5$\sigma$ discovery potential threshold, can be used to trigger follow-up observations with CTAO. This strategy maximizes the detection probability of sources producing both high-energy neutrinos and gamma rays. In this study, we evaluate CTAO’s ability to detect gamma-ray emission from sources preselected by KM3NeT based on their predicted neutrino flux, using simulated source populations and realistic instrument response functions.

\section{Simulation Methodology}

\subsection{Neutrino Source Populations with FIRESONG}
\noindent
Source populations are simulated using the open-source \texttt{FIRESONG} code (FIRst Extragalactic Simulation Of Neutrinos and Gamma-rays)~\cite{tung2021firesong}, which models extragalactic neutrino emitters as standard candles distributed according to a $\Lambda$CDM cosmology. The sources are assumed to be steady and follow a redshift evolution that traces the star formation rate (SFR). Key parameters such as the source density $\rho_{loc}$ and intrinsic neutrino luminosity $L_{\nu}$ are configurable, and in this study we restrict the intrinsic neutrino luminosity to values below $10^{52} \ \mathrm{erg} \ \mathrm{yr}^{-1}$, to avoid including extreme scenarios with unphysical energies. The total observed astrophysical neutrino flux is attributed to the ensemble of these sources. Each simulation yields a catalog of sources with assigned redshift, sky position, and expected neutrino flux, serving as input for subsequent gamma-ray detectability analysis. The spectral energy distribution of the neutrino flux is modeled as a power law, normalized at 100~TeV:

\[
E^2 \frac{dN}{dE} = A_\nu \left( \frac{E}{100\ \mathrm{TeV}} \right)^{\Gamma - 2},
\]

\noindent
where $A_\nu$ is the normalization constant of the neutrino flux at 100~TeV, and $\Gamma$ is the spectral index. This form ensures a flat $E^2 dN/dE$ representation for $\Gamma = 2$, consistent with Fermi acceleration expectations.

\subsection{Neutrino Detection Filtering with KM3NeT}
\noindent
To identify sources potentially detectable by the KM3NeT neutrino telescope, we apply a filter based on its discovery potential. This criterion defines the minimum neutrino flux necessary to achieve a $5\sigma$ detection within a given observation time, reflecting the instrument's ability to distinguish astrophysical sources above the background. The filter accounts for the effective area, directional acceptance, and expected background rate of the telescope for track-like events, and is applied using the latest KM3NeT sensitivity curves~\cite{vaneeden2024arca}. Sources passing this selection are considered viable neutrino detections and are then forwarded to the gamma-ray analysis stage.

\subsection{Gamma-ray Emission and EBL Absorption}
\noindent
For each selected source, a point-like morphology is assumed for the spatial component, while the gamma ray spectrum is described by a power law with exponential cutoffs at both low and high energies, following the parameterization:

\begin{equation}
    \frac{dN_\gamma}{dE} = A_{\nu} \left( \frac{E}{100\,\mathrm{TeV}} \right)^{-\Gamma} \exp\left( -\frac{E'_L}{(1+z)E} - \frac{E(1+z)}{E'_H} \right),
\end{equation}

\noindent
where $\Gamma$ is the spectral index, $A_{\nu}$ is the normalization related to the neutrino flux, and $E'_L$ and $E'_H$ are the low- and high-energy cut-off points in the source rest frame, respectively. The spectral shape follows the phenomenological model by Fiorillo et al.~\cite{fiorillo2021txs}, based on Halzen et al.~\cite{halzen2019txs}’s interpretation of the TXS0506+056 flare. Gamma-ray attenuation due to the Extragalactic Background Light (EBL) is included via the optical depth parametrization of Domínguez et al.~\cite{dominguez2011ebl}, applying an energy- and redshift-dependent absorption factor $\exp[-\tau(E,z)]$ that strongly suppresses flux from distant sources at multi-TeV energies.

\subsection{CTAO Simulation Setup and Detection Criteria}
\begin{figure}[h]
\centering
\includegraphics[width=0.6\textwidth]{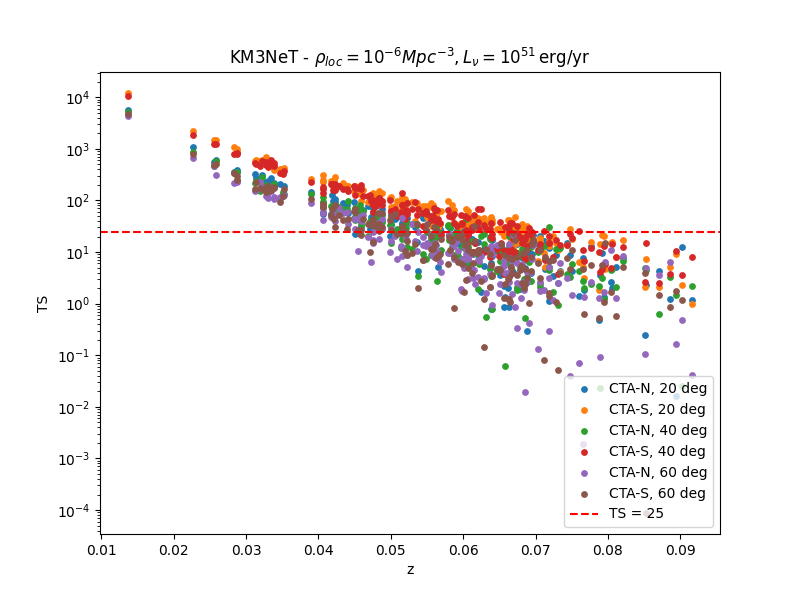}
\caption{Test Statistic (TS) as a function of redshift for a simulated source population with fixed  neutrino luminosity $L_{\nu}$ of $10^{51}\ \mathrm{erg\,yr^{-1}}$ and source density $\rho_{loc}$ of $10^{-6}\ \mathrm{Mpc}^{-3}$. The horizontal dashed line marks the detection threshold at $\mathrm{TS}=25$.}

\label{fig:ts_redshift}
\end{figure}

\noindent
The gamma-ray detectability of each source is assessed using the \texttt{ctools}~\cite{knodlseder2016gamma} and \texttt{GammaLib} software, which simulates event lists and background, and performs maximum likelihood analyzes. The simulations employ the \texttt{prod5-v0.1} instrument response functions for the CTAO Alpha configuration. Each source is observed for 30 minutes at a zenith angle corresponding to its culmination position. A detection is claimed if the Test Statistic (TS), defined as twice the log-likelihood ratio between the null (background-only) and alternative (signal + background) models, satisfies $\mathrm{TS} \geq 25$, roughly equivalent to a $5\sigma$ significance under Wilks' theorem. To include observational geometry and geomagnetic effects, performance is evaluated at both CTAO sites (North and South), for zenith angles of $20^\circ$, $40^\circ$, and $60^\circ$. For each zenith bin, simulations are run for three magnetic field alignments relative to the telescope pointing azimuth:  North (N) yields lower detectability due to stronger field-induced shower broadening, South (S) gives higher efficiency, and Average (AV) represents a directionally averaged case. Figure~\ref{fig:ts_redshift} shows the typical TS dependence on redshift for a simulated source population with fixed neutrino luminosity $L_{\nu}$ $10^{51}\ \mathrm{erg\ yr}^{-1}$ and source density $\rho_{loc}$ $10^{-6}\ \mathrm{Mpc}^{-3}$. The dashed line at $\mathrm{TS}=25$ marks the detection threshold for the 30-minute exposure.

\section{Results}

\subsection{Performance Comparison: CTAO-North vs. CTAO-South}
\begin{figure}[htbp]
  \centering
  \begin{subfigure}[b]{0.8\textwidth}
    \centering
    \includegraphics[width=\textwidth]{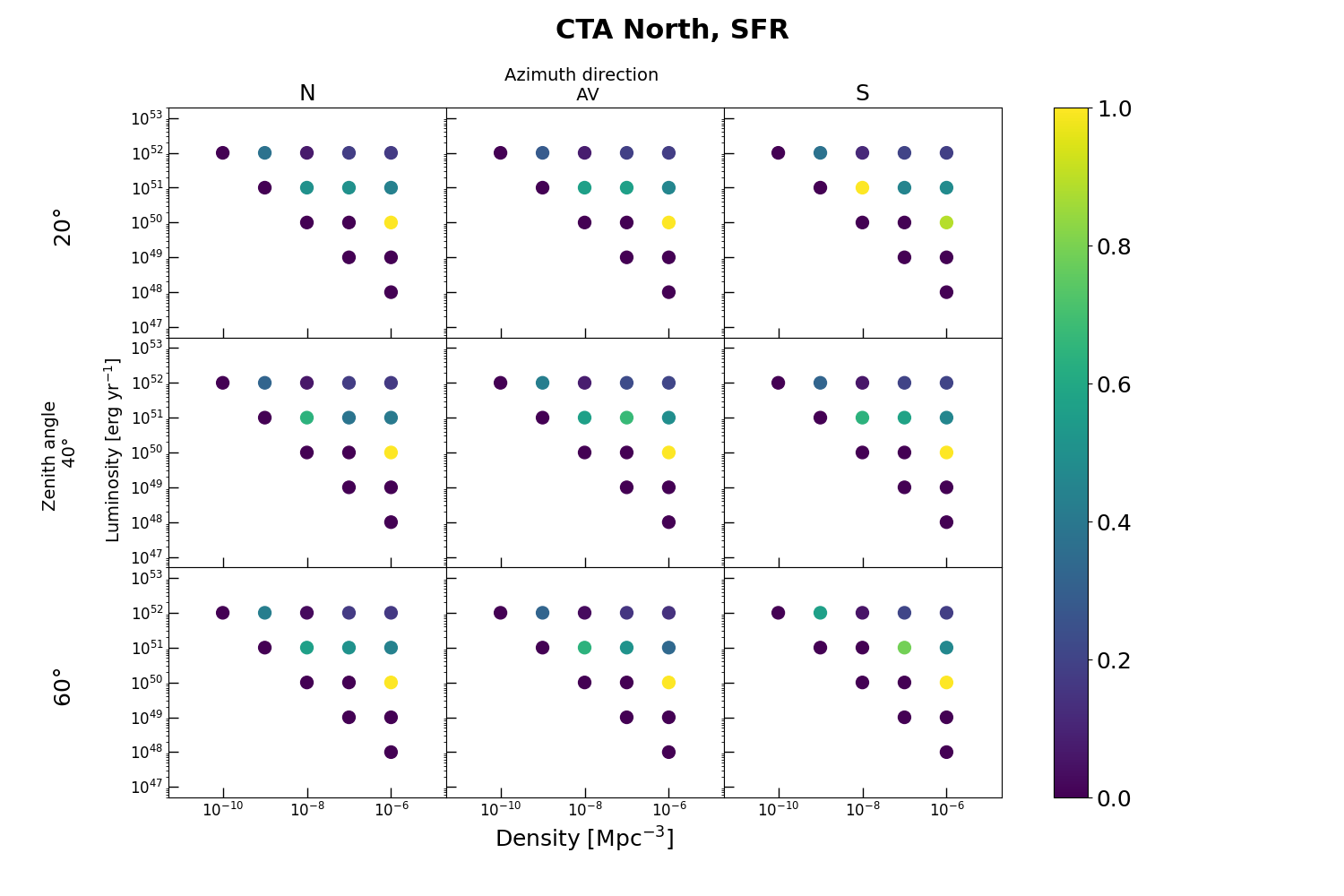}
    \caption{CTAO-North}
    \label{fig:ctao_north}
  \end{subfigure}
  \vspace{0.5em}
  \begin{subfigure}[b]{0.8\textwidth}
    \centering
    \includegraphics[width=\textwidth]{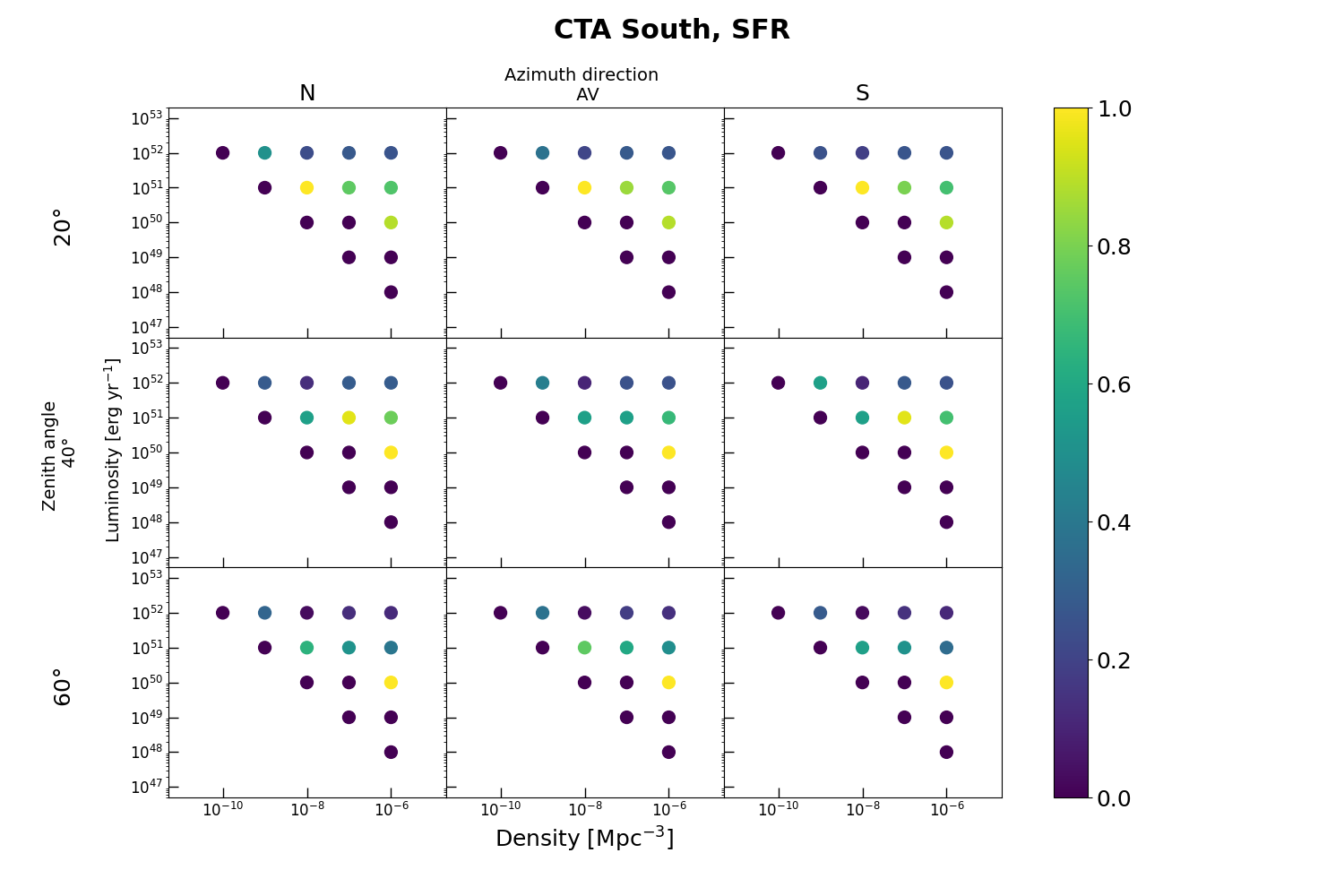}
    \caption{CTAO-South}
    \label{fig:ctao_south}
  \end{subfigure}
  \caption{Detection probability maps for CTAO-North (top) and CTAO-South (bottom), for sources with luminosities \textbf{below} \(10^{52} \ \mathrm{erg} \ \mathrm{yr}^{-1}\) and source densities ranging from \(10^{-11}\) to \(10^{-6}\)~Mpc\(^{-3}\). The sources are simulated to follow the evolutionary model of the star formation rate (SFR).}
  \label{fig:CTAO_prob}
\end{figure}

\noindent
Figure~\ref{fig:CTAO_prob} shows the detection probability maps for CTAO-North and CTAO-South. The southern site (CTAO-South) will benefit significantly from the proposed CTA+ upgrade, which includes the addition of two LSTs and five SSTs. In the current Alpha configuration, the absence of LSTs in the southern array reduces the sensitivity below 100 GeV. The analysis focuses on sources with luminosity below \(10^{52} \ \mathrm{erg} \ \mathrm{yr}^{-1}\). Rows correspond to different zenith angles (20°, 40°, and 60°), while columns represent magnetic field alignments relative to the telescope pointing azimuth: North (N), South (S) and Average (AV). 

\subsection{Detection Efficiency}
\begin{figure}[h]
\centering
\includegraphics[width=0.6\textwidth]{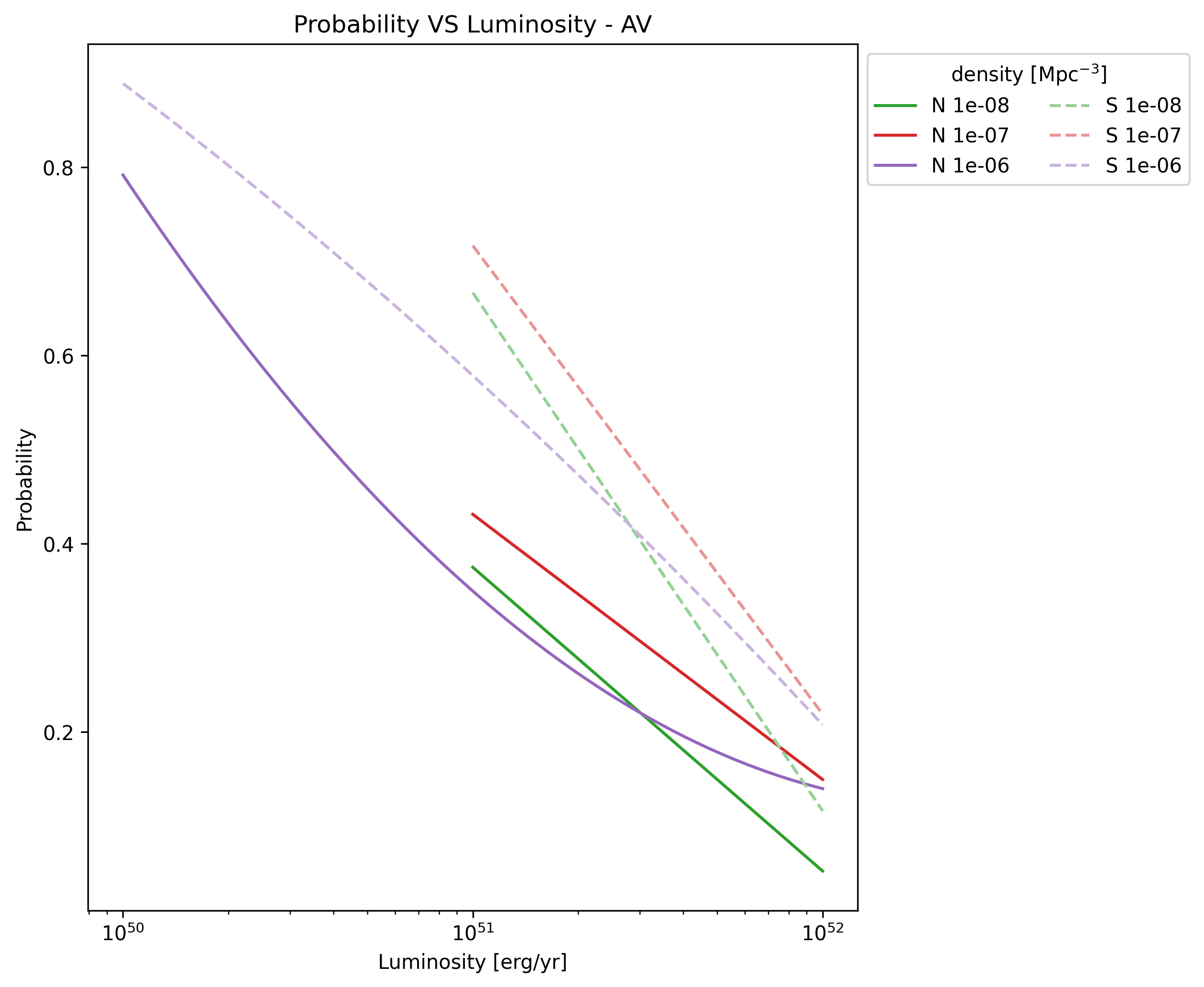}
\caption{Detection probability as a function of source luminosity for the AV configuration, limited to sources with luminosity below \(10^{52} \ \mathrm{erg} \ \mathrm{yr}^{-1}\). The southern configuration (S) generally shows higher detection probabilities due to KM3NeT’s advantageous Mediterranean location.}
\label{fig:prob_lum}
\end{figure}

\noindent
The detection efficiency as a function of source luminosity has been evaluated for the AV configuration, which offers intermediate behavior. KM3NeT maintains a robust detection capability throughout the analyzed luminosity range. This reflects the efficiency of the telescope not only in detecting highly luminous sources, as might be expected, but also in identifying relatively weak emitters. This sensitivity is critical, as many extragalactic neutrino sources may have moderate or low luminosity. As a result, the ability of KM3NeT to detect these less intense but scientifically significant sources substantially expands the potential for discoveries in neutrino astronomy. Figure~\ref{fig:prob_lum} shows the AV results as an example; the southern configuration performs best due to KM3NeT’s favorable location in the Mediterranean Sea.

\section{Discussion}
\noindent
The results presented in this study highlight the potential of a coordinated multi-messenger strategy involving KM3NeT and CTAO to identifying extragalactic neutrino sources. KM3NeT, with its ARCA detector located in the Mediterranean Sea, is particularly sensitive to track-like neutrino events from the southern sky. This complements the capabilities of CTAO, especially its southern site in Paranal, Chile, which is crucial for follow-up gamma-ray observations of many promising neutrino emitters. Compared to IceCube, which has a broader field of view for Northern Hemisphere sources but lower angular resolution, KM3NeT offers improved pointing precision, allowing more targeted follow-up observations. Although IceCube has already provided evidence for possible neutrino associations with blazars (e.g., TXS~0506+056~\cite{icecube2018science}), location and angular performance of KM3NeT make it better suited to resolve and identify fainter or more distant sources in the southern sky. One critical limitation revealed by our simulations is the current absence of Large-Sized Telescopes (LSTs) in the CTAO-South “Alpha” configuration. Since LSTs are essential for low-energy sensitivity (20–150~GeV), their absence significantly reduces the ability to detect extragalactic sources whose gamma-ray flux is softened by EBL absorption. The proposed CTA+ upgrade, which will add two LSTs and five SSTs at the southern site~\cite{ctaplus2023inaf}, is expected to lower the energy threshold and improve sensitivity, particularly for sources with faint or high redshift. These results suggest that follow-up strategies should be tailored by site: CTAO-North for nearby, high-luminosity sources, and CTAO-South (post-upgrade) for southern targets. Real-time alerts from KM3NeT combined with CTAO’s rapid-response capabilities will maximize the scientific return from both transient and steady sources.

\section{Conclusion and Outlook}
\noindent
We presented a performance study of the Cherenkov Telescope Array Observatory (CTAO) in detecting gamma-ray emission from steady extragalactic sources preselected by the KM3NeT neutrino telescope. Using realistic source simulations with \texttt{FIRESONG} and gamma-ray follow-up with \texttt{ctools} and \texttt{GammaLib}, we evaluated CTAO’s detection prospects in its “Alpha” configuration. Our results show that CTAO can detect a subset of KM3NeT-identified neutrino sources, especially those at low redshift and moderate gamma-ray luminosity. CTAO-North currently performs better due to the presence of LSTs, while the CTA+ upgrade at the southern site will significantly improve sensitivity at low energies. The complementarity of KM3NeT and CTAO—combining precise neutrino localization with rapid gamma-ray follow-up—emerges as a powerful strategy for source identification. Future work will include simulations with CTA+ response functions, incorporate transient sources, and explore a wider luminosity range beyond $10^{52} \ \mathrm{erg} \ \mathrm{yr}^{-1}$. Coordinated observations between KM3NeT and CTAO will be key to uncovering the sources of cosmic neutrinos and probing the most powerful accelerators in the Universe.

\section*{Acknowledgements}
\noindent
This research made use of ctools, a community-developed analysis package for Imaging Air Cherenkov Telescope data. ctools is based on GammaLib, a community-developed toolbox for the scientific analysis of astronomical gamma-ray data.  This research has made use of the CTA instrument response functions provided by the CTA Consortium and Observatory, see https://www.ctaobservatory.org/science/cta-performance/ (version prod5-v1) for more details. G.M.C. acknowledges financial support from the European Union—Next Generation EU RFF M4C2 under the project IR0000012—CTA+ (CUP C53C22000430006), announcement N.3264 on 28/12/2021: "Rafforzamento e creazione di IR nell'ambito del Piano Nazionale di Ripresa e Resilienza (PNRR)."


\begin{thebibliography}{99}



\bibitem{hofmann2024cherenkov}
W.~Hofmann and R.~Zanin,
\emph{The Cherenkov Telescope Array},
in \emph{Handbook of X-ray and Gamma-ray Astrophysics},
Springer (2024), pp.~2787--2833.

\bibitem{km3netoverview}
S. Adrián-Martínez et al. (KM3NeT Collaboration),  
“Letter of intent for KM3NeT 2.0,”  
\emph{Journal of Physics G: Nuclear and Particle Physics},  
vol. 43, no. 8, 084001, 2016.  
\url{https://doi.org/10.1088/0954-3899/43/8/084001}

\bibitem{adrian2016letter}
S.~Adrian-Martinez \textit{et al.}, ``Letter of intent for KM3NeT 2.0,'' \emph{Journal of Physics G: Nuclear and Particle Physics}, vol.~43, no.~8, p.~084001, 2016.

\bibitem{icecube}
IceCube Collaboration,  
\emph{The IceCube Neutrino Observatory},  
\url{https://icecube.wisc.edu} [Accessed: June 2025].

\bibitem{tung2021firesong}
C.~F.~Tung, A.~Kheirandish, A.~Achterberg, M.~Ahlers, and A.~Keivani,
\emph{FIRESONG: A python package to simulate populations of extragalactic neutrino sources},
J. Open Source Softw. \textbf{6} (2021) no.61, 3194.

\bibitem{vaneeden2024arca}
T.~J.~van Eeden, M.~D.~Filipović, and the KM3NeT Collaboration,
\emph{Astronomy potential of KM3NeT/ARCA230},
PoS \textbf{ICRC2023} (2024) 1075.

\bibitem{halzen2019txs}
F.~Halzen, A.~Kheirandish, T.~Weisgarber, and S.~P.~Wakely,
\emph{A model for the multi-messenger emission of TXS~0506+056},
Astrophys.\ J.\ Lett.\ \textbf{874} (2019) L9.


\bibitem{fiorillo2021txs}
D.~F.~G.~Fiorillo, K.~Satalecka, I.~Taboada, and C.~F.~Tung,
\emph{Time-dependent modeling of the 2014–2015 neutrino flare from TXS~0506+056},
Astrophys.\ J.\ \textbf{917} (2021) 70.


\bibitem{dominguez2011ebl}
A.~Domínguez et al.,
\emph{Extragalactic background light inferred from AEGIS galaxy-SED-type fractions},
Mon.\ Not.\ Roy.\ Astron.\ Soc.\ \textbf{410} (2011) 2556--2578.

\bibitem{knodlseder2016gamma}
J.~Knödlseder et al.,
\emph{GammaLib and ctools: A software framework for the analysis of astronomical gamma-ray data},
Astron.\ Astrophys.\ \textbf{593} (2016) A1.

\bibitem{ctaplus2023inaf}
INAF,
\emph{Progetto CTA+ – Rafforzamento e creazione di Infrastrutture di Ricerca (IR)},
\url{https://pnrr.inaf.it/progetto-ctaplus/} [Accessed: June 2025].

\bibitem{icecube2018science}
IceCube Collaboration, Multi-messenger observations of a flaring blazar coincident with high-energy neutrino IceCube-170922A, \textit{Science} \textbf{361}, eaat1378 (2018), doi:10.1126/science.aat1378.


\end{thebibliography}
\end{document}